\documentclass[12pt]{iopart}
\usepackage[english]{babel} 
\usepackage[english]{babel}
\usepackage{latexsym}
\usepackage{graphicx}
\usepackage{listings}
\usepackage[applemac]{inputenc}
\usepackage{caption}
\usepackage{sidecap}
\usepackage{ifpdf}
\usepackage{iopams}
\usepackage{amsthm}
\usepackage{amssymb}
\usepackage{multirow}
\usepackage{epstopdf}
\usepackage{hyperref}
\usepackage{subfigure}
\usepackage{cite}

 \newcommand {\beq}{\begin{displaymath} }
 \newcommand {\eeq}{\end{displaymath}}
 \newcommand{\ben}{\begin{equation}}
 \newcommand{\een}{\end{equation}}

\begin{document}

\title[Identification of vortexes obstructing the dynamo mechanism]{Identification of vortexes obstructing the dynamo mechanism in laboratory experiments}

\author{A Limone$^1$, D R Hatch$^{1,3}$, C B Forest$^2$ and F Jenko$^1$}

\address{$^1$ Max Planck Institute for Plasma Physics, EURATOM Association, Boltzmannstr. 2, 85748 Garching bei M\"{u}nchen, Germany}
\address{$^2$ Department of Physics, University of Wisconsin, Madison, 1150 University Avenue, Madison, WI 53706, USA}
\address{$^3$ Institute for Fusion Studies, University of Texas at Austin, Austin, Texas 78712, USA}

\ead{angl@ipp.mpg.de}
\begin{abstract}
The magnetohydrodynamic dynamo effect explains the generation of self-sustained magnetic fields in electrically conducting flows, especially in geo- and astrophysical environments. Yet the details of this mechanism are still unknown, e.g., how and to which extent the geometry, the fluid topology, the forcing mechanism and the turbulence can have a negative effect on this process. We report on numerical simulations carried out in spherical geometry, analyzing the predicted velocity flow with the so-called Singular Value Decomposition, a powerful technique that allows us to precisely identify vortexes in the flow which would be difficult to characterize with conventional spectral methods. We then quantify the contribution of these vortexes to the growth rate of the magnetic energy in the system. We identify an axisymmetric vortex, whose rotational direction changes periodically in time, and whose dynamics are decoupled from those of the large scale background flow, is detrimental for the dynamo effect. A comparison with experiments is carried out, showing that similar dynamics were observed in cylindrical geometry. These previously unexpected eddies, which impede the dynamo effect, offer an explanation for the experimental difficulties in attaining a dynamo in spherical geometry.
\end{abstract}
\maketitle

\section{Introduction}

The magnetohydrodynamic (MHD) dynamo effect is considered to be the main mechanism responsible for the generation of self-sustained magnetic fields in geo- and astrophysical objects: The kinetic energy of an electrically conducting fluid is transformed into magnetic energy. According to this theory, a magnetic field can be generated from a electrically conducting fluid (e.g., liquid metal or plasma) when, in the governing equation - called the ``magnetic induction equation'' - the advection term dominates the diffusion one, at least sufficiently in order to amplify initial magnetic perturbations. The mathematical aspects of dynamo theory are well described in \cite{Moffatt1978,Bullard1954}, whereas the important role that this theory plays in astrophysics can be grasped in \cite{Dormy2007,Ruediger2004}. Experimental realizations of dynamo action in laboratory experiments have been successfully accomplished in the past (see, e.g., Refs. \cite{Gailitis2008} for the history and the results of the Riga experiment, or \cite{Stieglitz2001} for the Karlsruhe experiment). Numerical simulations of dynamo action are particularly important in the design phase of an experiment, and the seminal paper \cite{Dudley1989} provides a good example of such numerical studies.

The identification of flow geometries and topologies that are optimal for the attainment of a magnetic instability is crucial \cite{Covas1999, Moffatt1992}. In order to achieve the dynamo mechanism, experiments are focused on the investigation of the optimal setup which enhances the probability of achieving dynamo action. The design of the geometry, the choice of the fluid, the forcing mechanism, as well as the boundary conditions and their material (for plasma dynamo experiments, see \cite{Khalzov}) play a crucial role in creating a flow field whose geometry provides the right feedback mechanism for the creation of a self-sustaining magnetic field. However, the design of the geometry of an experiment is not enough for such a task because one should take into account also the detrimental effects that turbulence has on the dynamo process, as recent experimental investigations \cite{Kaplan2011,Ponty2007b} have shown. In \cite{Kian} it is experimentally observed that turbulence enhances the magnetic diffusivity of the liquid, an effect that hinders dynamo action. The effects of turbulence have also been investigated via numerical simulations: A fundamental work is described in \cite{Laval2006}, which studied whether turbulence (in a Taylor-Green flow and in periodic boxes) raises or lowers the dynamo threshold (i.e., the so-called critical magnetic Reynolds number $Rm_c$, above which the magnetic field growth takes place). The main result of \cite{Laval2006} is that the addition to the mean velocity field of a large-scale stochastic noise significantly increases the threshold, whereas for small-scale noise, the results depend on the correlation time of the noise and the magnetic Prandtl number (the ratio of the magnetic to the fluid Reynolds number). The results of our work also reinforce this conclusion by demonstrating that large scale vortices with specific features hinder the dynamo mechanism also in a bounded domain whose geometry is interesting for experimental purposes.

The aim of this work is to study in detail the dynamics of the flow in order to understand which configurations are detrimental for the dynamo process and which ones are favorable. The idea is the following: Decomposing the field via the Bullard-Gellman decomposition \cite{Bullard1954}, we identify the modes which turn out to have an impact on the dynamo threshold $Rm_c$. In order to facilitate this analysis, we analyze the velocity field - simulated by the DYNAMO code \cite{Bayliss2007, Reuter2008} - with a technique known as Singular Value Decomposition (SVD). This technique has been widely applied to the analysis of turbulent flows~\cite{berkooz} and, more recently, to a variety of transport processes in plasma microturbulence, including impurity transport~\cite{futatani}, electromagnetic transport~\cite{hatch12}, resistive ballooning plasma fluctuations~\cite{Beyer}, and saturation mechanisms~\cite{hatch11}. A pedagogical introduction to this decomposition and its applications can be found in \cite{linearalgebra}. 

The remainder of this paper is structured in the following way: In Section \ref{dtsg} a general background about dynamo theory in spherical geometry will be given, with focus on numerical and laboratory experiments; in Section \ref{svd_sec} we will briefly describe the Singular Value Decomposition, with emphasis on the decomposition of the flow fields predicted by the DYNAMO code; in Section \ref{water} we focus on the results of a ``water'' experiment (setting the magnetic conductivity to zero), showing the detailed structure and dynamics of the turbulent field found via SVD; in Section \ref{curva} we use the analysis of our results in order to explain in more detail the dependence of the dynamo threshold $Rm_c$ on turbulence (i.e., the curve $Rm_c(Re)$, previously found for this spherical system); Section \ref{strategies} deals with possible ways to control large scale turbulence in order to facilitate dynamo action.

\section{Dynamo experiments and simulations in spherical geometry}\label{dtsg}

\subsection{The Madison Dynamo Experiment}
\begin{figure}[!hbp] 
\includegraphics[width=6cm]{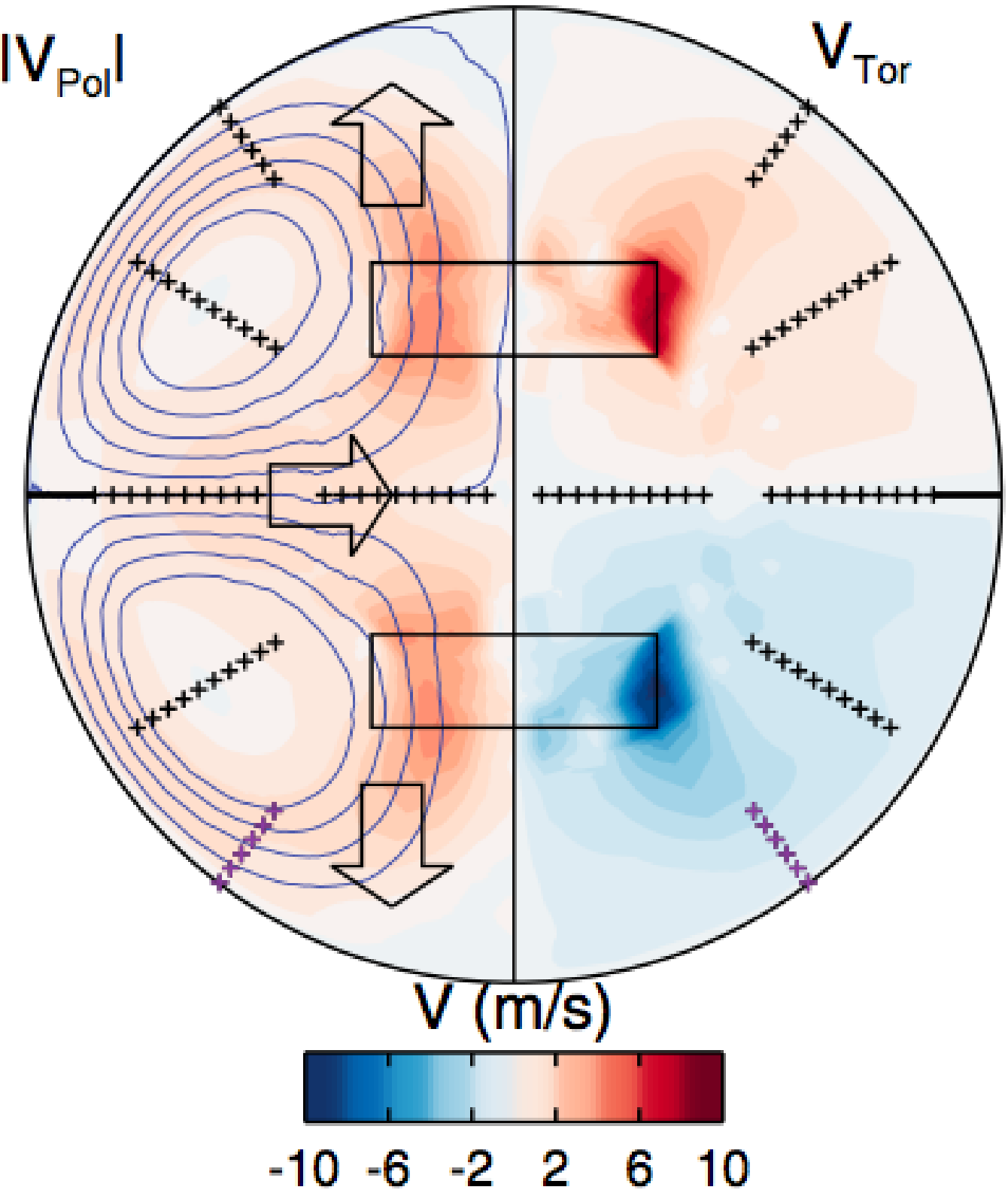}\centering\caption{An example of the mean flow of the MDE (cross section, $yz$-plane) computed using FLUENT$\textsuperscript{\textregistered}$. The hollow black boxes represent the impellers. The internal probe array is indicated by the black and purple crosses. Picture taken from \cite{Kaplan2011}.} \label{mde_flow}
 \end{figure}

The Madison Dynamo Experiment (MDE) consists of a spherical vessel (radius of 0.533 m) of stainless steel containing liquid sodium. The flow is stirred by two counter-rotating impellers of 30.5 cm diameter that enter the sphere through each pole (a cutaway view of the experiment can be found - for instance - in \cite{Spence2006}), having a topology similar to the Von K\'arm\'an Sodium experiment \cite{Berhanu2007, Monchaux2007, Ravelet2008}.  The impellers thrust fluid outwards to the poles. In each hemisphere, the mean flow streams along the walls to the equatorial plane, where it rolls back in towards the center of the sphere (Fig. \ref{mde_flow}), creating a so-called \emph{s2t2} flow, as studied by Dudley and James in \cite{Dudley1989}. In addition to this poloidal circulation, the two flow cells counter-rotate in toroidal direction, as imposed by the sense of rotation of the propellers. Driving is provided by two 75 kW motors. The radial component of the magnetic field is measured by an array of temperature-compensated Hall probes mounted to the sphere’s surface. Magnetic fields within the sphere are measured by linear arrays of Hall probes inserted into the sodium within stainless steel sheaths. 

Finally, two external electromagnets, in a Helmholtz configuration coaxial with the impellers, apply a nearly uniform magnetic field throughout the sphere. The MDE is directly inspired by the work of Dudley and James, who performed numerical kinematic dynamo studies and found this particular two-cell flow geometry to be among the most efficient simple spherical dynamos. They showed that this flow minimizes the threshold $Rm_c$ of the dynamo instability.

The experiment achieves a magnetic Reynolds number of $Rm\sim100$ which implies a Reynolds number of $Re=\mathcal{O}(10^7)$ due to the fixed small magnetic Prandtl number $Pm = Rm/Re = \mathcal{O}(10^{-5})$ of liquid sodium. The current experimental setup, however, does not reach the dynamo threshold due to detrimental effects of turbulence, its critical magnetic Reynolds number $Rm_c$ being larger than $\sim100$ by some unknown factor.

\subsection{The DYNAMO code}

The DYNAMO code \cite{Bayliss2007} solves - in spherical geometry - the nonlinear equations of incompressible MHD, i.e., the \emph{magnetic induction equation} coupled with the Navier-Stokes equation, by computing three-dimensional time-dependent solutions for the velocity and magnetic fields. The simulated system has been designed to model the MDE, having the same geometrical and dynamical features. In particular, the code focuses on the so-called \emph{s2t2} type of flow (see \cite{Dudley1989} for a definition), although it could in principle study different forcing mechanisms (i.e., different flows). The code exploits a pseudo-spectral method based on spherical harmonics decomposition, and was originally developed by R. A. Bayliss at the University of Wisconsin-Madison \cite{BaylissPhD, Bayliss2007}, and afterwards extended and parallelized by K. Reuter \cite{Reuter2008}. The parallelization was performed by dividing the sphere in spherical shells. Further details on the physical model, the stirring force, the numerical setup and the parallelization of the code can be found in \cite{Reuter2008}. 

One of the previous key results obtained by the DYNAMO code was the discovery of a hydrodynamic instability in the weakly turbulent regime, i.e., the exponential growth of modes with $m = 2$ symmetry located in the shear layer region between the forced regions and the walls of the sphere \cite{Reuter2009b}. In the nonlinear saturation regime, these modes manifest themselves as hydrodynamic waves propagating in the zonal direction, oppositely directed in each hemisphere. Increasing the Reynolds number above the wave-dominated regime causes new bifurcations, finally leading to inhomogeneous anisotropic turbulence. Kinematic dynamo simulations using the time averaged flow or different snapshots of the velocity field did not exhibit dynamo action, whereas growing magnetic fields were found when considering the time-dependent flow. This result is linked to the presence of perpetual \emph{non-normal growth} due to the mathematical property of the magnetic induction equation’s linear operator, being time dependent and non self-adjoint  \cite{Tilgner2008}. Finally, the stability curve $Rm_c(Re)$ of this system, i.e., the dependence of the dynamo threshold on the fluid Reynolds number, was also determined \cite{Reuter2011}.

\subsection{Model}
\def\sgn{\mathop{\rm sgn}}

Liquid metals used in dynamo experiments can be modeled by using incompressible MHD, i.e., the \emph{extended} Navier-Stokes equation, which -- in its nondimensional form -- reads

\ben \frac{\partial \mathbf{v}}{\partial t}+Rm(\mathbf{v}\cdot\nabla)\mathbf{v}=-Rm\nabla P + Pm \nabla^2\mathbf{v}+Rm (\nabla\times\mathbf{B})\times\mathbf{B}+Rm\mathbf{F},\een coupled with the (nondimensional) \emph{magnetic induction equation}
\begin{equation}\frac{\partial \mathbf{B}}{\partial t}=Rm\nabla\times(\mathbf{v}\times\mathbf{B})+\nabla^2\mathbf{B},\label{inductionequation}\end{equation} with the constraints \ben\nabla\cdot\mathbf{v}=0,\een and \ben\nabla\cdot\mathbf{B}=0,\een where $Rm=\mu_0 \sigma RU$ is the magnetic Reynolds, $R$ the characteristic length associated with the system (the radius of the sphere, in this case), $U$ the characteristic velocity and $\sigma$ the conductivity of the medium, $Re=RU/\nu$ is the fluid Reynolds number, $\nu$ the viscosity of the fluid, $\mathbf{j}=\mu_0^{-1}\nabla\times\mathbf{B}$ is the current density, $p$ the pressure, $\mathbf{F}$ the forcing term. We assume a constant mass density $\rho=1$. In order to correctly achieve the nondimensional form of the incompressible MHD system, time has been scaled using the resistive diffusion time $\tau_\sigma=\mu_0 \sigma R^2$. As stated above, in the case $Rm\gg1$, the advection term dominates and a dynamo mechanism takes place if the geometry of the flow $\mathbf{v}$ can support it. Following a standard convention, the characteristic velocity $U$ is calculated as the time-averaged rms velocity $U=\overline{\sqrt{\langle |\mathbf{v}|^2\rangle}}$, where the overbar represents temporal and the the brackets $\langle\cdot\rangle$ denote spatial averaging. The time averaging is performed during the kinematic phase of a run, i.e. when the back-reaction of the magnetic field on the flow is negligible. The driving term is implemented as a constant body force $\mathbf{F}$, designed in order to produce a flow with the same topology as that of the MDE. This flow consists of two counter-rotating cells (see, e.g., \cite{Bayliss2007}), and -- due to the dominant spectral components of the resulting velocity field, the flow is known as \emph{s2t2}. Because of its axisymmetric feature, the driving term is best described in a centered, cylindrical coordinate system $(\rho,\phi,z)$, where the $z$-axis corresponds to the axis of symmetry of the MDE impellers (the $\theta = 0$ axis in spherical coordinates). The body force hence reads

\ben F_{\rho}=0,\quad F_{\phi}=\epsilon \sgn(z)\rho^3 r^{-3}_d\sin{\frac{\pi \rho}{2 r_d}}+\gamma, \een \ben F_z=(1-\epsilon)\sgn(z)\sin{\frac{\pi\rho}{r_d}}+\delta,\een with $\epsilon=0.1$, $\gamma=0.05$ and $\delta=0.3$, as also described in \cite{Reuter2009a}.
On the boundary of the fluid region, velocity field are subjected to no-slip boundary conditions, whereas the boundary condition on the magnetic field is derived based on the assumption that the space outside the spherical fluid conductor is free from electrical currents (electrically insulating vessel). The assumption is justified since the wall of stainless steel of the MDE is a significantly worse conductor than the volume of liquid sodium in the inside \cite{BaylissPhD}. 

\section{Singular Value Decomposition and application to the DYNAMO data} \label{svd_sec}
SVD is a mathematical technique used in the framework of multivariate statistics in order to simplify the representation of datasets, or in order to reduce the memory which is needed to store them, by reducing the number of variables used to describe the data. This reduction is obtained by substituting the original variables with a set of new variables which are less numerous, via an orthogonal linear transformation of the former variables, ordering the new set of variables according to their ``information'' quantity: The first variables contain more information than the others and describe the large scale features of the geometrical distribution; the other variables add less information to the description, focusing on the small-scale features. It is worth noticing that the expressions \emph{large scale} and \emph{small-scale} are used here in a context which is -- in general -- different from the context of turbulence in fluids; nevertheless, if the SVD analysis is applied to fluid mechanics data, the large (small) scale features of the distribution of the data often coincide with the large (small) scale features of the turbulent fields (see, e.g., \cite{futatani}). A pedagogical introduction can be found in textbook \cite{linearalgebra} or in Ref. \cite{WallRecht}. In this last reference, several applications of this decomposition can be found and a similar matrix formalism is employed. Guidelines about how to implement a SVD routine can be found in the classical textbook \cite{numericalrecipes}.

With this new representation, it is possible to reduce the complexity of the dataset by projecting the data on the first few new variables and limiting oneself to analyzing only the information described by these first principal components. In other words, the technique suggests an optimal set of eigenmodes that capture the main features of the data, by tailoring the eigenmodes to the dataset in an optimal way.  By means of this method it is easier to understand the behavior of the system by isolating the most important dynamics. In this study, the method is used to decompose the velocity fields in a more general (and appropriate) way, facilitating a detailed study of the dynamo mechanism, and shedding light on the properties of the transfer of the energy between the kinetic and the magnetic components of the system.


\subsection{Bullard-Gellman decomposition}
In the following, the Bullard-Gellman decomposition (also called \emph{poloidal-toroidal} or \emph{Mie} decomposition) will be extensively used, since it is particularly advantageous in spherical geometry. The incompressible velocity field $\mathbf{v}(\mathbf{x})$ is divergence-free, hence it can be written as 

\begin{equation} \mathbf{v}(\mathbf{x})=\nabla \times \nabla  \times [s(\mathbf{x})\mathbf{x}] + \nabla\times[t(\mathbf{x})\mathbf{x}],\label{torpol}\end{equation} where the scalars $s(\mathbf{r})=s(r,\theta,\phi)$ and $t(\mathbf{r})=t(r, \theta, \phi)$ are called - respectively - the poloidal and toroidal stream functions. These functions can be - in turn - expanded in terms of spherical harmonics, i.e.,
\begin{equation} s(r,\theta,\phi)=\sum^{\infty}_{l=0}\sum^l_{m=0}s(r)^m_l \mathcal{Y}^m_l(\theta,\phi),\end{equation}
\begin{equation} t(r,\theta,\phi)=\sum^{\infty}_{l=0}\sum^l_{m=0}t(r)^m_l \mathcal{Y}^m_l(\theta,\phi),\end{equation} where $\mathcal{Y}^m_l$ is the spherical harmonics with ``angular momentum'' and azimuthal wavenumbers $l$ and $m$. The azimuthal wavenumber is limited to $m\geq 0$ since all vector fields are real-valued. Using the triangular truncation, the total number of relevant spectral modes is given by \begin{equation}(l_{max}+1)(l_{max}+2)/2-1,\end{equation} where $l_{max}$ is the maximum ``angular momentum wavenumber''. 
The monopole mode $l = 0$ is not required due to the divergence free constraints. To account for dealiasing, the upper third of the spectrum is truncated by choosing $\frac{2}{3}N_\theta -1 \geq l_{max}$ and $N_\phi = 2N_{theta}$,  where $N_\theta$ and $N_\phi$ are the number of latitudinal and longitudinal points used in the real space representations of the vector fields \cite{ReuterPhD, Reuter2008}.

\subsection{SVD decomposition of the velocity field}\label{SVD_math}  

In this section we describe in details how we used SVD on our datasets and with which consequences. 
The DYNAMO data, which are analyzed via SVD, are the velocity (or magnetic) fields calculated by the simulations. Two ways of application of the SVD are possible. The first one is in spatial representation, i.e., the information contained in a vector field $\mathbf{V}(\mathbf{r}_i,t_j$), where $\mathbf{r}_i$ is one of the gridpoints of the simulation and $t_j$ is the $j$-th time step, can be rearranged in a 2D matrix $A(i,j)$ where the row index spans the spatial grid \emph{and} the three components of the field $\mathbf{V}$, while the column index $j$ spans \emph{only} the temporal dimension. The second way is in the spectral representation, i.e., the information contained in a vector field is described by the poloidal and toroidal radial profiles as a function of the radial distance from the center of the sphere and the mode numbers $l$ and $m$. In the spectral representation, we indicate as $\mathbf{r}_i$ the generalized spatial point $(r,l,m)$ and $t_j$ is again the $j$-th time step. The data can be rearranged in a 2D matrix $A(i,j)$ where the row index spans the generalized spatial grid and the two components of the field while the column index $j$ spans the temporal dimension. This rearranging of the field 
can be easily accomplished by ``unfolding'' the spatial domain into a 1D array with $i=1,2,...,N_{space}$, where $N_{space}$ is, in the spatial representation, equal to $3n_r n_{\theta} n_{\phi}$ (the prefactor 3 due to the three dimensionality of the original field) or to $2 n_r n_{modes}$, where the prefactor 2 due to the presence of a toroidal and a poloidal components, $n_{modes}$ is the maximum number of relevant modes of the simulation, i.e. $(l_{max}+1)(l_{max}+2)/2-1$, where $l_{max}$ is the maximum ``angular momentum wavenumber'' (see \cite{ReuterPhD}). We chose to analyze the fields in spectral representation, since it provides datasets that occupy less disk space and the SVD routines are consequently faster.

Based on this idea, we represent the spatiotemporal DYNAMO data as the $A \in \mathbb{C}^{ N_{space} \times N_t}$ matrix, where $N_t$ is the number of time steps considered, as also shown in \cite{benkadda} in a fusion plasma context.  With this new arrangement of the data in a 2D matrix, it is possible to produce the singular value decomposition (which we accomplish using a parallelized SVD solver~\cite{scalapack}), thus decomposing matrix $A$ as

\ben A_{ij}=\sum^{N_{SVD}}_{k=1}\sigma_k u_k(\mathbf{r}_i)v_k(t_j),\label{SVD}\een where we indicated with $N_{SVD}$ the value $\min (N_{space}, N_t)$. In general, the two most important properties of the SVD are the optimality and orthonormality of the $N_{SVD}$ modes.  For our purposes, however, optimality plays the most important role: It guarantees that a truncation of the decomposition,  
\ben A^{(n)}_{ij}=\sum^{n<N_{SVD}}_{k=1}\sigma_k u_k(\mathbf{r}_i)v_k(t_j),\label{SVD}\een produces a smaller error than any other possible decomposition of the same rank.  SVD has the property that it minimizes the Euclidean distance $|| A-~A^{(n)}||$ between the original dataset $A$ and its truncated version $A^{(n)}$, where $ || A || =~ \sqrt{\sum_{ij}A_{ij}^2}$ is the Frobenius norm, i.e., it minimizes the truncation error of the compressed version of the data. This feature is mathematically equivalent to the maximization of the variance of the dataset along the first new generalized reference axes.  In the context of the current analysis, this means that each SVD mode captures more of the dynamics than any other possible representation.  The orthonormality property states that, \beq\sum^{N_{SVD}}_{i=1} u_m (\mathbf{r}_i) u_n (\mathbf{r}_i) = \sum^{N_{SVD}}_{j=1} v_m (t_j) v_n (t_j)=\delta_{mn}. \eeq  It is important to note, however, that this orthonormality condition is lost when the data (in spectral representation) are transformed back into vector fields in spatial representation, hence the scalar product is not equivalent to the energy scalar product, but it is closely related to it. The singular values $\sigma_k$ are ordered in descending order of magnitude: $\sigma_1>\sigma_2>...>\sigma_{N_{SVD}}$. The value $\sigma_k$ quantifies the relative weight of the $k$-mode, i.e., the amount of information contained in the $k$-mode. Using the jargon of spectral analysis, the spatial fields $u_k (\mathbf{t}_i)$ are the new generalized basis functions and the temporal functions $\sigma_k v_k(t_j)$ can be seen as the spectrum of the data as a function of the time. In the following, $u_k(\mathbf{r}_i)$ are called the SVD modes. If the singular values $\sigma_k$ decay fast as a function of $k$, as it typically happens, most of the information is contained in the first few modes, and the data can be successfully compressed, without a great loss of quality, by truncating the sum in Eq. \ref{SVD} at $k=r$, with $r<N_{SVD}$. In the following, we will indicate with $A^{(r)}$ this truncation of the original dataset.

Another statistical property of SVD is the possibility of quantifying the information entropy of the decomposed datasets. The dimensionless value
\ben p_k =\frac{ \sigma^2_k} {\sum^{N_{SVD}}_{k=1} \sigma^2_k}\label{pik}\een  measures the weight of the $k$-th mode, and it is limited to $[0,1]$. Moreover \beq\sum^{N_{SVD}}_{k=1} p_k=1.\eeq Thus, it behaves as a probability distribution function, to which it is possible to apply the formula of the normalized information entropy $H$, in order to quantify the degree of order of the data, i.e.,

\ben H=- \frac{\sum^{N_{SVD}}_{k=1} p_k \log p_k}{\log N_{SVD}}.\label{entropy}\een
When the data are in a very ordered configuration, i.e., only one mode is needed to represent the whole dataset ($A_{ij}=\sigma_1 u_1 v_1$), then $p_k=\delta_{1k}$ (here, $\delta_{ij}$ is the Kronecker symbol), hence $H=0$. On the contrary, in the ``highly disordered'' configuration, when the information is uniformly distributed among all the SVD modes, the entropy value is maximum, i.e., $H=1$.

\section{SVD analysis of the virtual water experiment} \label{water}
In order to characterize quantitatively the hydrodynamics of the simulated system, as a first step we apply the SVD analysis to the data produced by a hydrodynamical simulation, i.e., setting to zero the conductivity of the medium, at numerical fluid Reynolds numbers $Re_0=\{600, 1100, 3000\}$. $Re_0$ is a parameter read by the DYNAMO code before a simulation is started; in order to obtain the real fluid Reynolds number $Re$ of a particular numerical simulation, $Re_0$ is multiplied with the characteristic velocity $U$ of the simulation \cite{ReuterPhD} because the characteristic velocity can only be determined \emph{a posteriori} via the equations $U=\overline{v_{rms}}$ and $\overline{v_{rms}}=\sqrt{\langle v^2\rangle}$ (the angle brackets denote averaging in space and the overline denotes averaging in time, which is performed during the quasi-stationary phase of the flow after the transient phase). The reason why $Re_0$ needs to be corrected lies in the fact that $Re$ must depend upon the fluid velocity, as its definition states.  In the simulations discussed in this paper, $U\sim 0.5$. The number of radial grid points in the sphere is $n_r = 512$, whereas the spectral resolution is $l_{max}=\{52, 52, 180\}$ respectively. We then analyze the flow when it has reached a saturated state, considering $N_t=\{1550, 700, 400\}$ snapshots of the velocity field (i.e., the number of time slices which constitute the columns of the matrix $A$, see section \ref{SVD_math}). The numbers $N_{t}$ are chosen according to the available computer resources. With these input parameters, the resulting $N_{SVD}$ turns out to be exactly $\{1550, 700, 400\}$ respectively. We will see that, among these $N_{SVD}$ modes, only a few are necessary in order to reconstruct the field without a great loss of quality. In the following, we will separately analyze in detail the first two modes.

\subsection{Singular values}\label{svalori}
As already stated, every singular value $\sigma_k$ quantifies the relative importance of the $k$-th mode. The singular values decay exponentially in the medium range of $k$-modes and even more steeply at small values of $k$. This eventuality is particularly favorable, since it has as a consequence that the information of the dynamics is condensed in the first SVD modes, whereas further modes do not significantly improve the understanding of the overall dynamics. Although this result seems to suggest that adding more modes (i.e., using the SVD tools on more time steps) does not add further details to the decomposition, it should be noted that using a larger number of time steps can also improve the statistics of the first modes, providing a mechanism for encapsulating the information even better in the first modes. Table \ref{tabellare} lists the percentage of information content of the first 6 modes calculated via the definition of $p_k$, Eq. \ref{pik} (i.e., the relative amount of information contained in the $k$-th mode) with different $Re_0$. As the table shows, at higher $Re_0$, the flow becomes progressively more turbulent and, comparatively, more energy is drained from the mean field and stored in the other extraneous modes.

\begin{table}[ht] 

\caption{Information content of the first 6 $k$ modes calculated via $p_k$.} 
\centering \begin{tabular}{ c c c c c c c } 
\hline\hline 
  & $k=1$&$k=2$&$k=3$&$k=4$&$k=5$&$k=6$   \\ \hline
$Re_0=600$ & 66.6 \%& 9.6\%& 5.4\%& 2.9\%& 2.4\%& 1.8\%\\
$Re_0=1100$ & 56.7 \%& 16.7\%& 5.1\%& 3.6\%& 2.6\%& 1.7\%\\
$Re_0=3000$ & 54.9 \%& 15.2\%& 5.3\%& 4.1\%& 2.5\%& 1.9\%\\
\hline
 \end{tabular}
 \label{tabellare}

 \end{table}

\subsection{First SVD mode}

\begin{figure} [!t]
 \begin{minipage}{.48\textwidth}\includegraphics[width=8cm, angle=0]{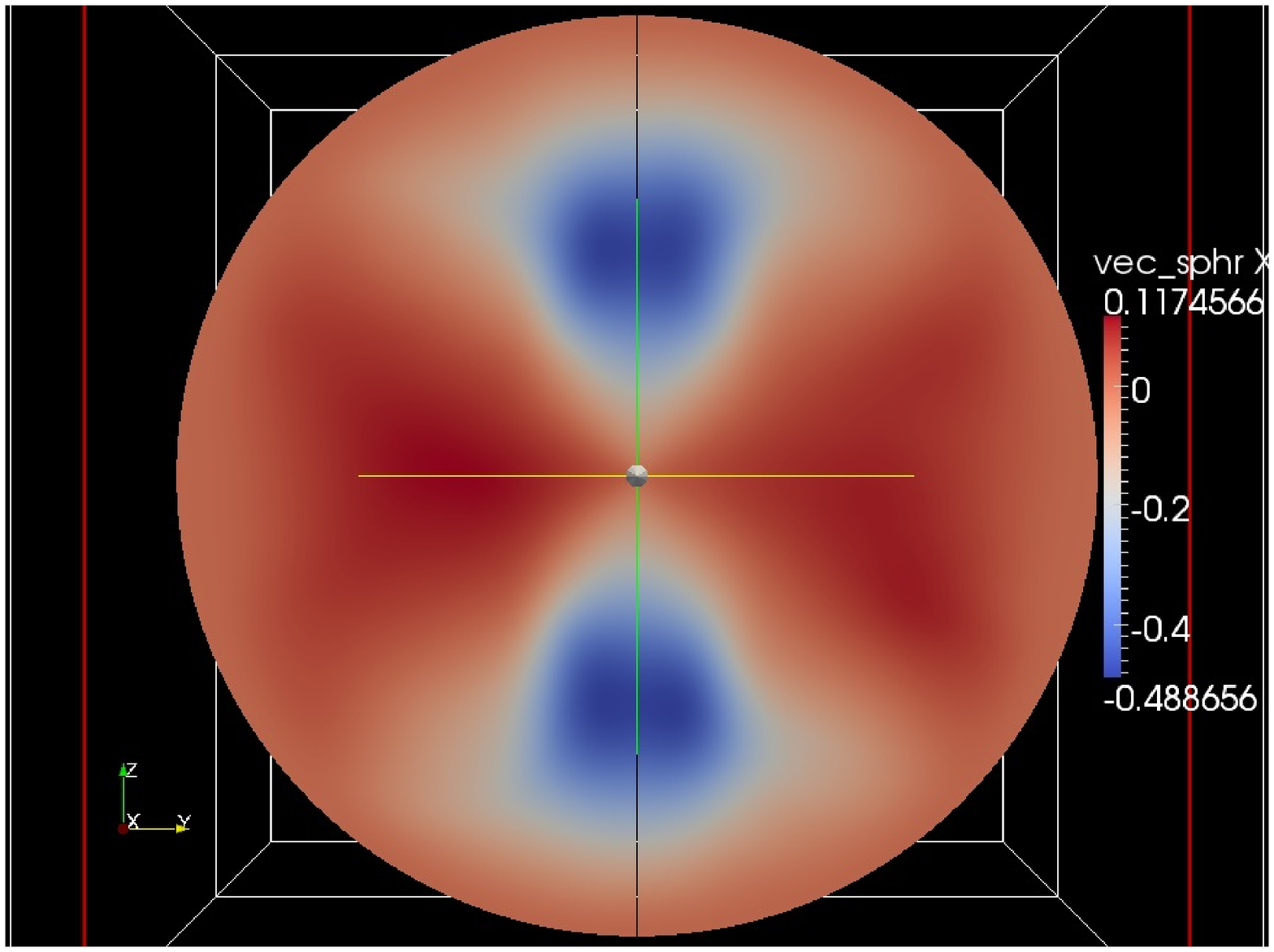}\centering\caption{First SVD eigenfunction, $Re_0=600$: The color represents the magnitude of the $r$-component of the field.} \label{1POD_vec_r}\end{minipage} 
  \hspace{15mm}
 \begin{minipage}{.48\textwidth}\includegraphics[width=8cm,angle=0]{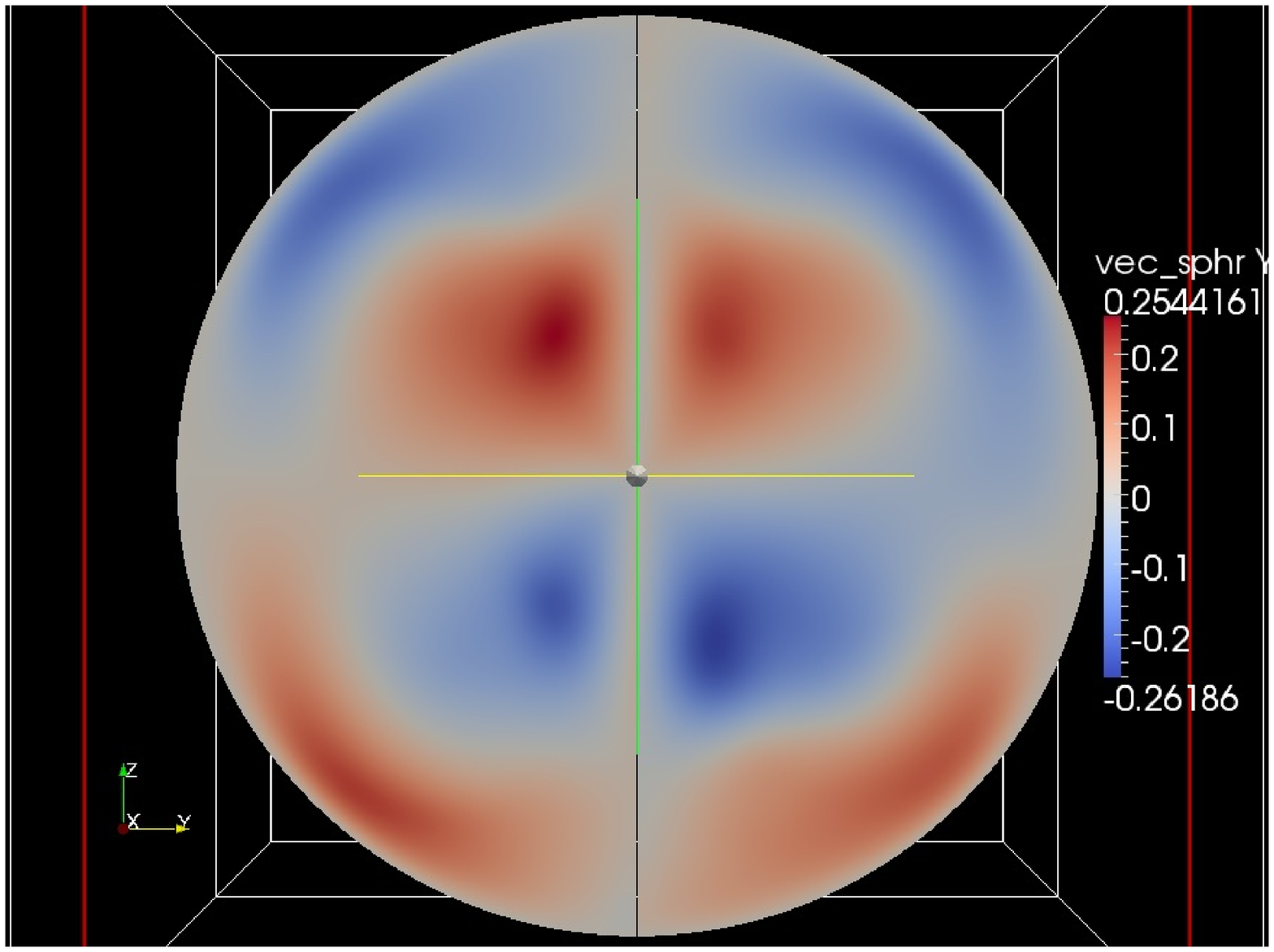}\centering\caption{First SVD eigenfunction, $Re_0=600$: The color represents the magnitude of the $\theta$-component of the field.} \label{1POD_vec_theta}\end{minipage} 
 \end{figure}

\begin{SCfigure}
 \begin{minipage}{.5\textwidth} \includegraphics[width=7.5cm]{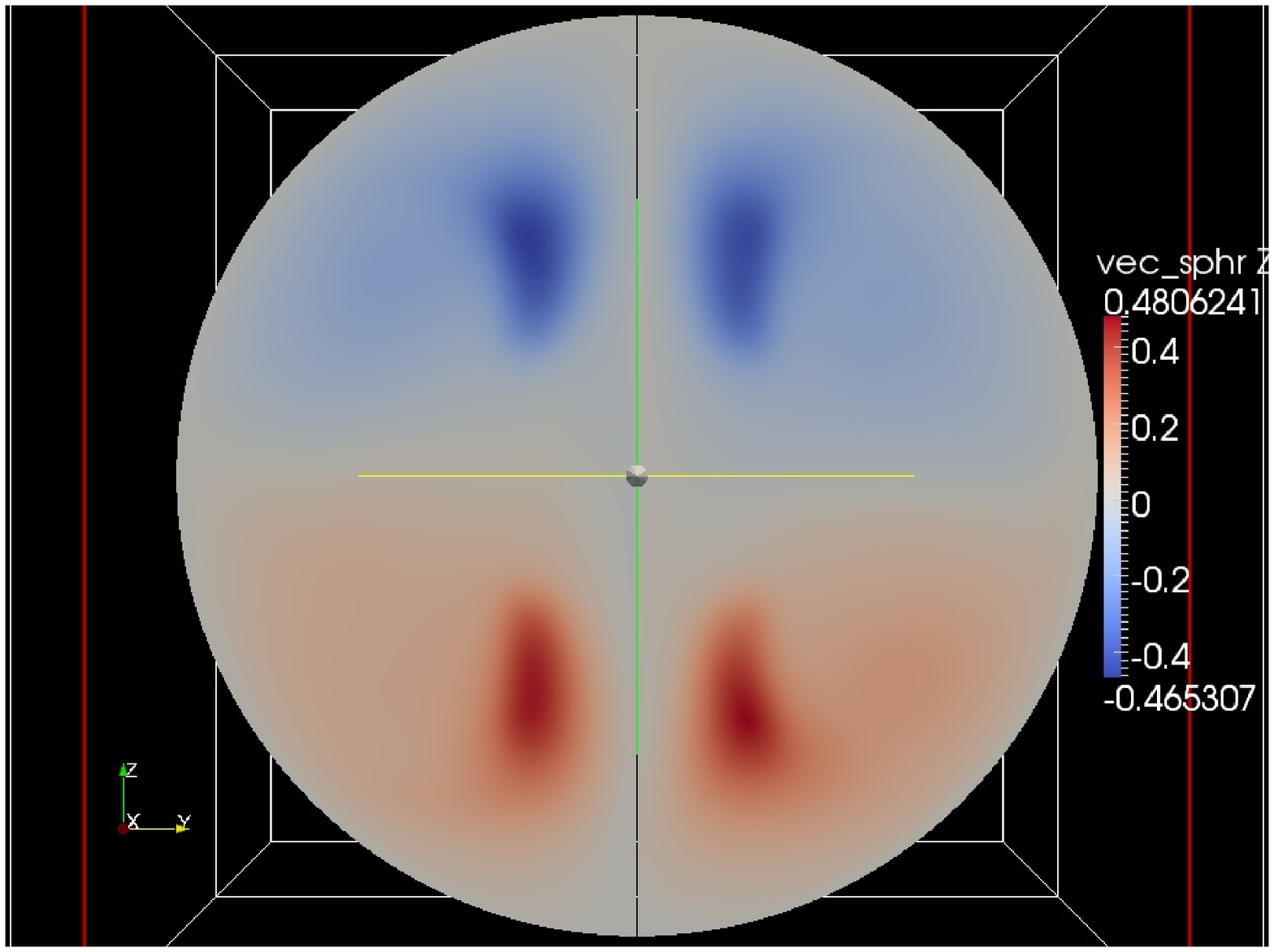}\centering\caption{First SVD eigenfunction, $Re_0=600$: The color represents the magnitude of the $\phi$-component of the field.} \label{1POD_vec_phi}\end{minipage}  
 \end{SCfigure}

\begin{figure}[!t]
 \begin{minipage}{.48\textwidth} \centering \includegraphics[width=5.7cm, angle=-90]{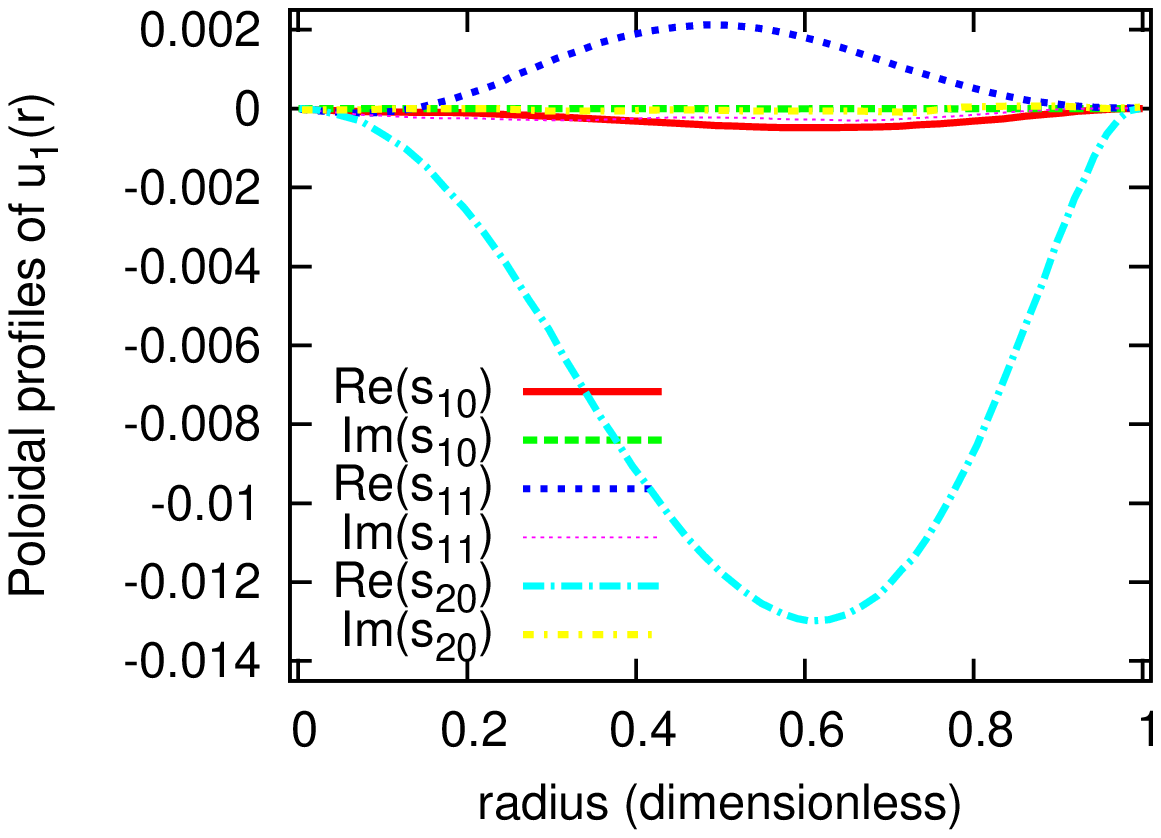}\centering\caption{Radial profile of the poloidal modes of the first SVD eigenfunction, $Re_0=600$.} \label{1POD_pol_new}\end{minipage}
 \hspace{5mm}
  \begin{minipage}{.48\textwidth}\centering \includegraphics[width=5.7cm,angle=-90]{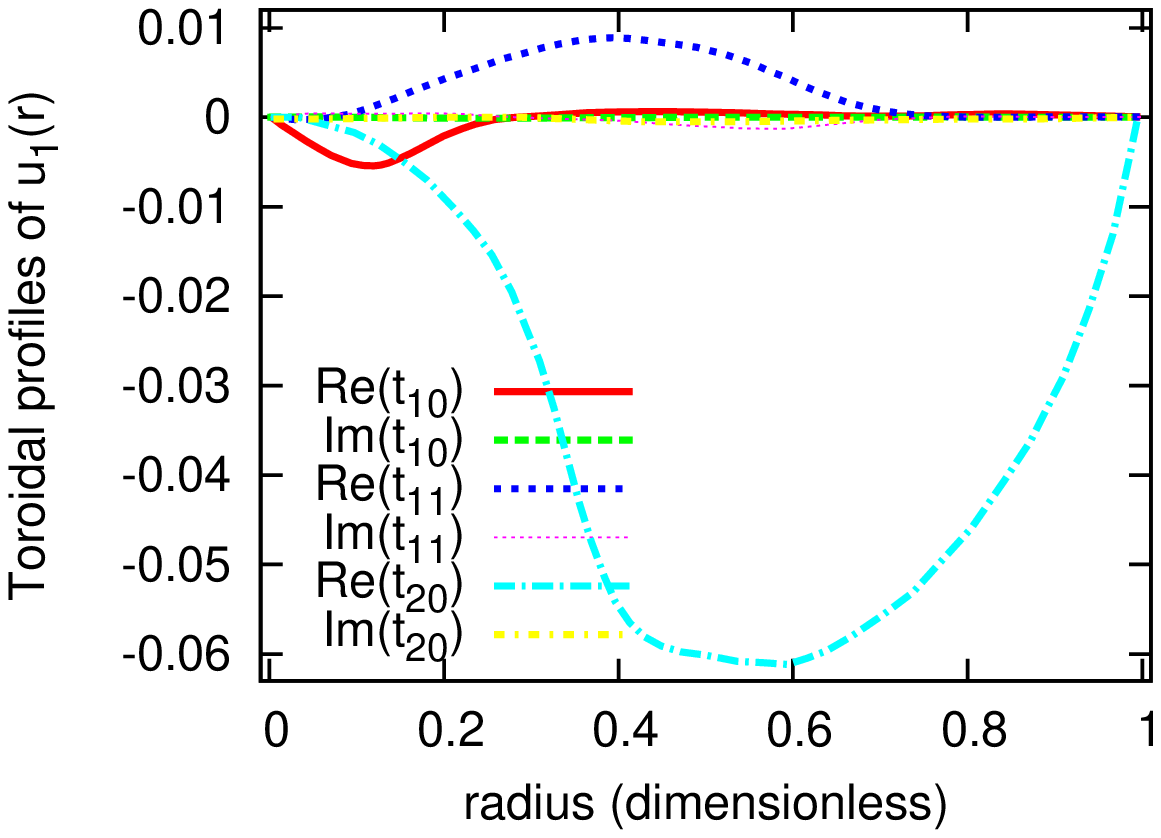}\centering\caption{Radial profile of the toroidal modes of the first SVD eigenfunction, $Re_0=600$.} \label{1POD_tor_new} \end{minipage}
 \end{figure}

We expect that the first SVD mode, i.e, $u_1(\mathbf{r})$ broadly captures the essential features of this hydrodynamical experiment. As mentioned above, it turns out that $u_1(\mathbf{r})$ is very similar to the mean field, which in turn resembles the Dudley \& James input flow, as the Figures \ref{1POD_vec_r}-\ref{1POD_vec_phi} show. These figures display a cross section of the sphere, i.e. the $yz$ plane; whereas their color shows - respectively - the radial, the $\theta$- and the $\phi$-components of the first SVD mode. As the color scales suggest, the flow speed is maximal in the vicinity of the impellers, and the typical counter-rotating nature of the motion can be seen thanks to the $\phi$-component plot. In other words, the shape of the original \emph{s2t2} flow can be recognized. A caveat is needed: In the original Dudley \& James configuration, the flow is driven outwards to the poles by the impellers and then it streams along the walls to the equatorial plane; on the contrary, Figure \ref{1POD_vec_r} shows that the first mode behaves oppositely. However, once this mode is multiplied by $\sigma_1 v_1$ in order to reconstruct the field, the original behavior is recovered since $\sigma_1 v_1<0$ (as Figure \ref{POD_time0001} shows; note that $\mathfrak{Im}  \{\sigma_1 v_1\}\sim 0$). We show the radial profiles of the spectral functions $s_{lm}(r)$ and $t_{lm}(r)$ at $Re_0=600$ in Figures \ref{1POD_pol_new}-\ref{1POD_tor_new}. These Figures suggest that the most important modes are still those of the \emph{s2t2} flow. It should be noted that poloidal stream functions have different physical dimensions from toroidal ones (see definition, Eq. \ref{torpol}), therefore direct comparisons of the scales of the plots in Figures \ref{1POD_pol_new}-\ref{1POD_tor_new} should be made only between stream function of the same kind (i.e., only between poloidal or only between toroidal ones). The situation is completely analogous at higher $Re_0$, just noisier, since the flow is more turbulent. 

The singular value associated with this mode is $\sigma_1=96.9$ at $Re_0=600$, which means that the information content of this first mode is $\sigma^2_1(\sum^{N_{SVD}}_{k=1}\sigma^2_k)^{-1}=0.66$ of the total.

\subsection{Second SVD mode}\label{secondSVDmode}
\begin{figure}[!t]
 \begin{minipage}{.48\textwidth} \centering \includegraphics[width=5.7cm, angle=-90]{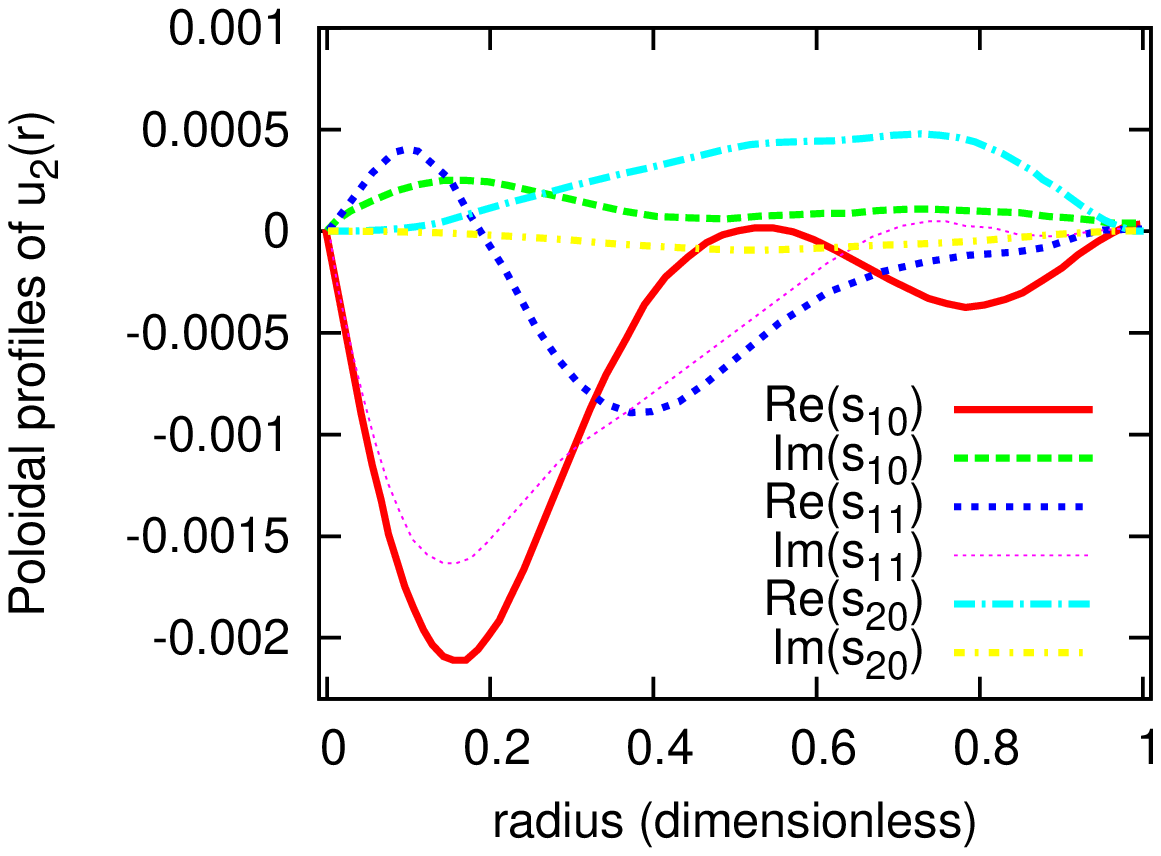}\centering\caption{Radial profile of the poloidal modes of the second SVD eigenfunction, $Re_0=600$.} \label{2POD_pol_new}\end{minipage}
 \hspace{5mm}
  \begin{minipage}{.48\textwidth}\centering \includegraphics[width=5.7cm,angle=-90]{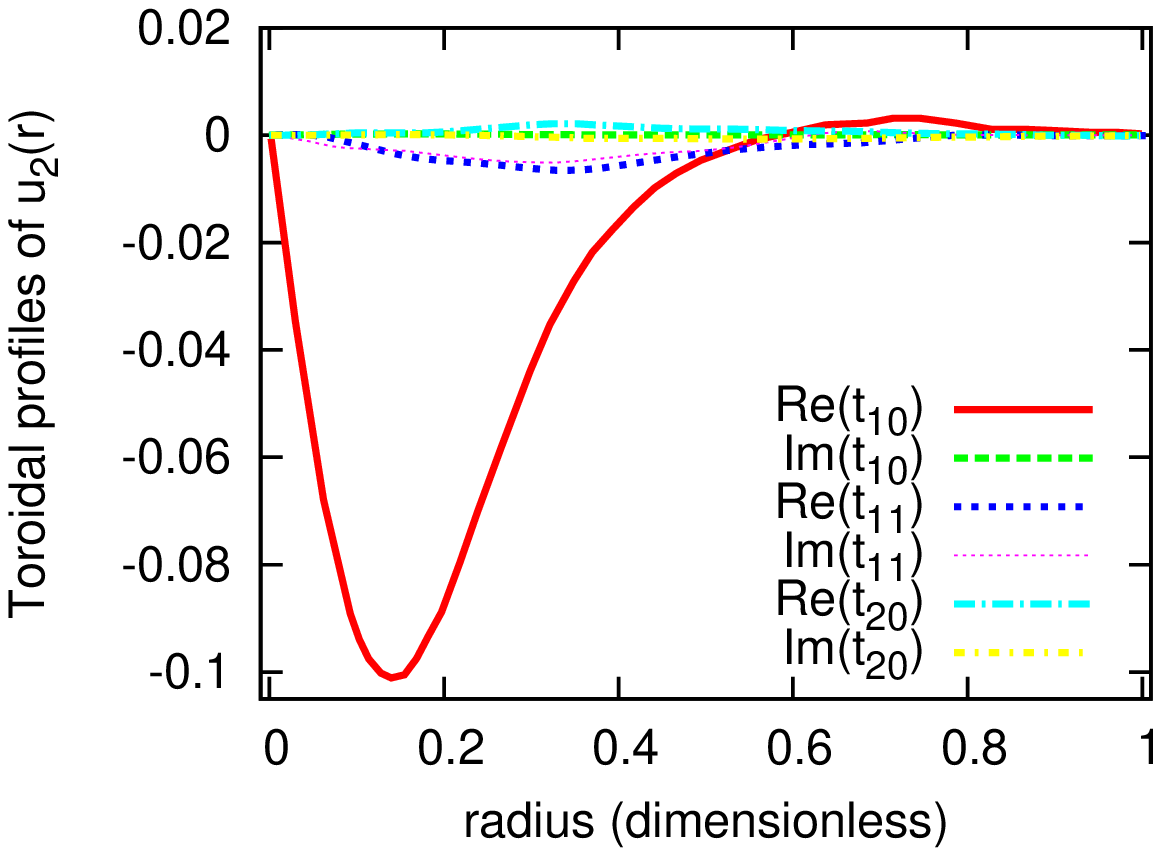}\centering\caption{Radial profile of the toroidal modes of the second SVD eigenfunction, $Re_0=600$.} \label{2POD_tor_new} \end{minipage}
 \end{figure}

\begin{figure}[!hbp] \includegraphics[width=12cm]{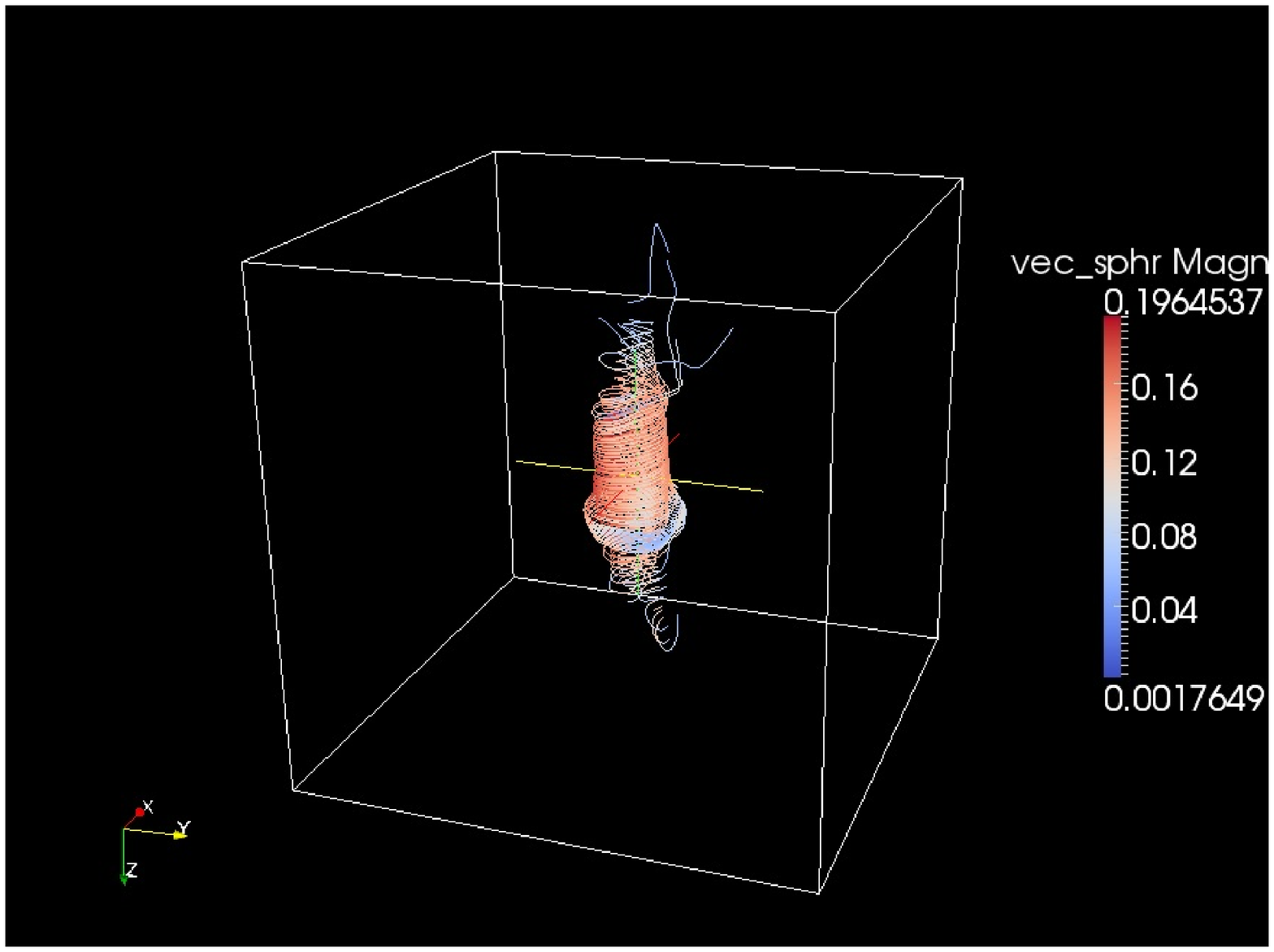}\centering\caption{Field lines of the second SVD eigenfunction, $Re_0=600$. The helical structure described above can be recognized. The color represents the magnitude of the field.} \label{2POD_mag}\end{figure}

The second SVD eigenfunction is also a significant feature of the dynamics. For instance, at $Re_0=600$ the associated singular value $\sigma_2=36.8$ is such that this mode contains alone 9.6\% of the total information (see Section \ref{svalori}). Let us analyze the spectral and spatial behavior of this mode. Figures \ref{2POD_pol_new}-\ref{2POD_tor_new} show the radial profiles of the poloidal and toroidal stream function at $Re_0=600$ (as stated above, the situation at higher $Re_0$ is analogous). As these plots suggest, $u_2(\mathbf{r})$ consists of three main components: (a) a toroidal vortex $t_{10}$ with a strong activity at $r\sim0.14$; (b) a poloidal circulation $s_{10}$ which provides the velocity field with a vertical component (i.e., a $z$-component); this component has a relatively strong amplitude at about the same distance as the toroidal vortex, making $u_2(\mathbf{r})$ helical; (c) noisy toroidal and poloidal components with negligible amplitudes. In other words, the vortex has a smooth and large scale spatial dependence, with an elongated helical shape oriented along the axis of symmetry of the forcing mechanism. It does not show any counter-rotating feature, as the \emph{s2t2} flow; on the contrary, its “winding” configuration is equal in the two hemispheres. Figure \ref{2POD_mag} shows a snapshot of the 3D fieldlines: The helical structure is readily recognized.

Note that this flow is characterized, at first glance, by a puzzling breaking of the symmetry expected from the forcing mechanism; the flow is driven by counter-rotating impellers, and yet $u_2(\mathbf{r})$ is characterized by a uniform rotation orientation on the axis between the impellers. The expected symmetry can only be retrieved - in an average sense - when one follows the temporal evolution of the mode, which exhibits alternating phases of opposite rotational orientation (see below).

\subsection{Temporal evolution}\label{temporal_ev}
\begin{figure}
\vspace{5mm}
   \begin{minipage}{.48\textwidth} \includegraphics[width=5.5cm, angle=270]{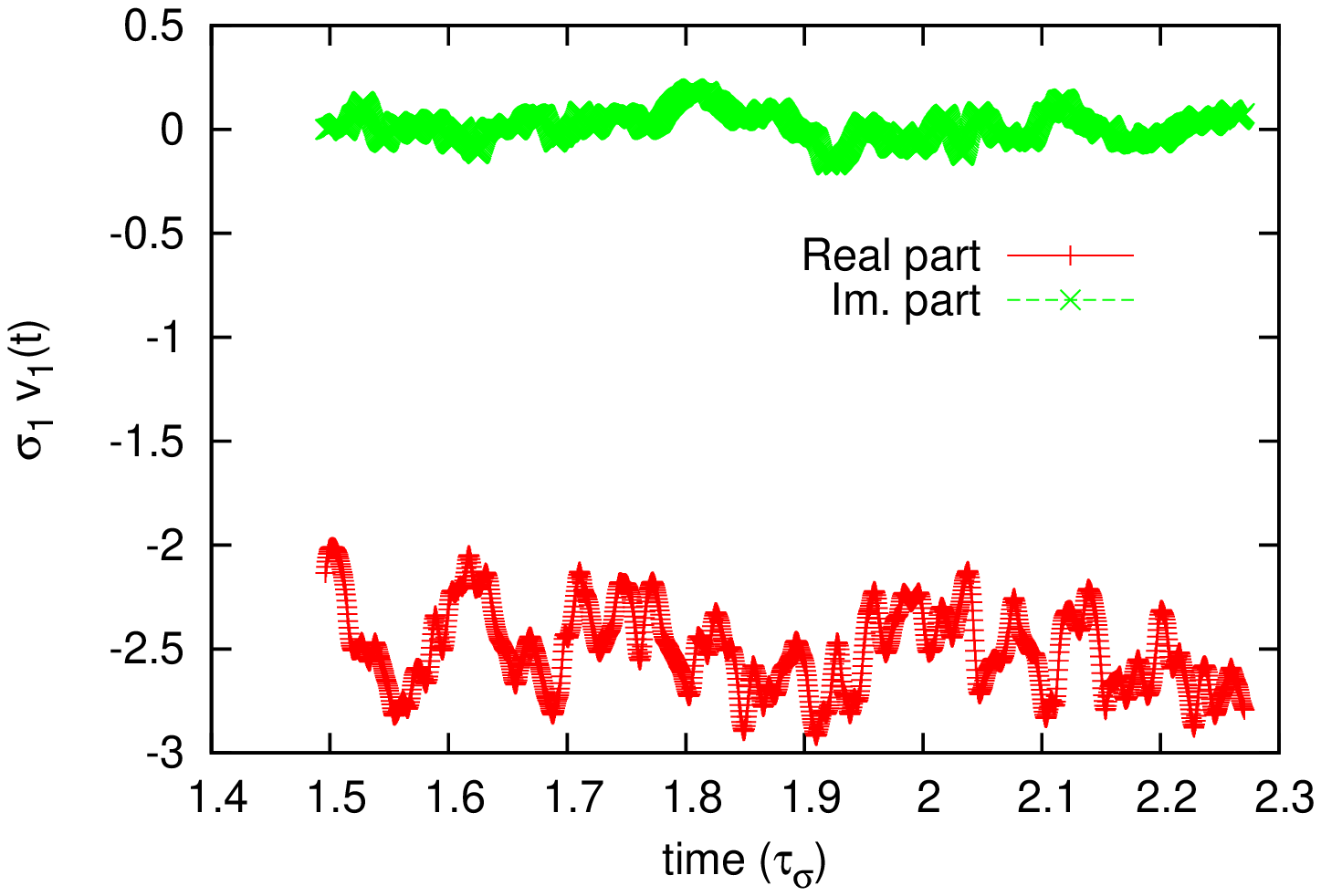}\centering\caption{Real and imaginary parts of the SVD temporal eigenfunction $\sigma_1 v_1(t)$, $Re_0=600$ \cite{MePre}.} \label{POD_time0001}\end{minipage}
   \hspace{10mm}  
   \begin{minipage}{.48\textwidth}\includegraphics[width=5.5cm, angle=270]{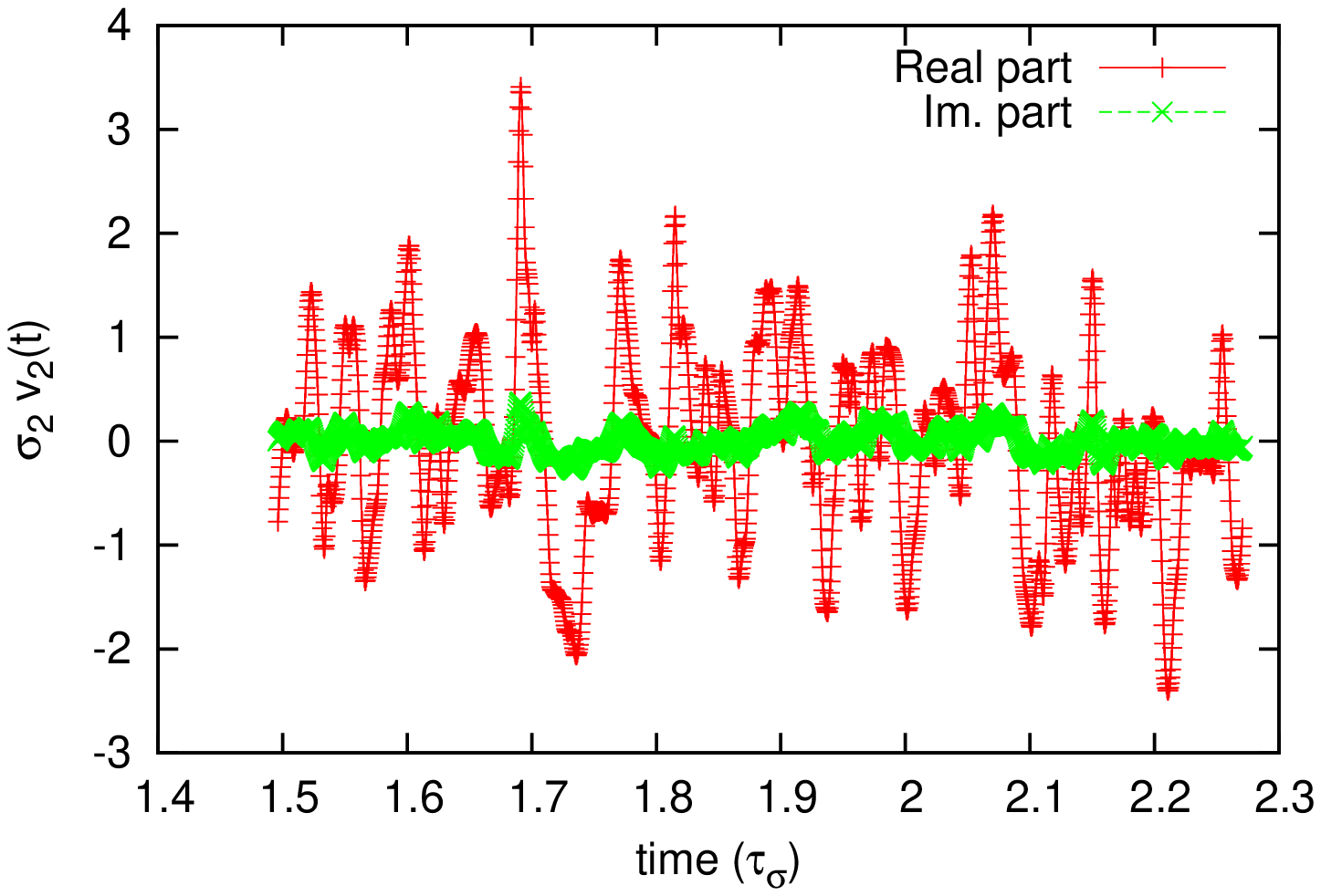}\centering\caption{Real and imaginary parts of the SVD temporal eigenfunction $\sigma_2 v_2(t)$, $Re_0=600$ \cite{MePre}.} \label{POD_time0002}\end{minipage}  
\end{figure}

The SVD analysis provides a detailed description of the dynamics of the $u_k(\mathbf{r})$ modes: The temporal eigenfunctions $\sigma_1 v_1(t)$ and $\sigma_2 v_2(t)$ describe the (decoupled) dynamics of the associated spatial modes. For instance, we show the results with $Re_0=600$ (at higher $Re_0$ the outcome does not change). As can be seen in Figures \ref{POD_time0001} and \ref{POD_time0002}, the imaginary parts of the two time series are negligible compared to the real parts. With regard to the first SVD mode, we note that the time series of the real part is an oscillation around a stable value (around -2.5, value which will be used in the next sections) which never crosses the $\sigma_1 v_1(t)=0$ axis. This means that, as we expect from the behavior of the \emph{s2t2} flow, the dynamics of the velocity field is basically an \emph{s2t2} flow with a superimposed oscillating variation of the local magnitude of the vector field, conserving the rotational direction in each hemisphere. The behavior of the $\sigma_2 v_2(t)$ is different: The real part changes sign, being limited roughly between $+1.5$ and $-1.5$ (values used in the next sections). In other words, the helical vortex alternates phases with opposite rotational directions.  This is translated, spatially, in a reversal of the vortex described in Section \ref{secondSVDmode}. These dynamics are difficult to extract without the use of SVD, which constructs basis functions specifically suited to the problem, and also clearly elucidates the temporal dynamics of each basis function. 

The dynamics of this hydrodynamic simulation can be summarized as follows. The usual counterrotating \emph{s2t2} flow is produced by the forcing mechanism and it constitutes the main component of the flow, namely a stationary background flow. This \emph{s2t2} flow is subjected instantaneously to increasing or decreasing activities around the mean configuration, as shown by the time trace in Figure \ref{POD_time0001}. Superimposed upon this background flow, a vortex-like component extends along the axis of symmetry, with alternating activity between two opposed configurations, one with the same rotation direction of the \emph{s2t2} flow in the northern hemisphere, the other one in the southern hemisphere. The additivity property of the decomposition states that the field associated with this secondary motion reinforces the magnitude of the velocity in the northern (southern) hemispheres and weakens the other one, when it is in the former (latter) configuration: In addition to the expected Dudley \& James background flow, each hemisphere has a pulsating activity near the $z$ axis, and this activity is in counter-phase with respect to the activity in the other hemisphere.

\subsection{Comparison with previous results}
In this section, we draw a comparison between our results and those obtained experimentally for a turbulent von K\'arm\'an swirling flow, where two impellers counterrotate inside a cylindrical cavity \cite{navarra, delarco}. For this experiment it was possible to reach fluid Reynolds numbers of the order of $10^4$. A broken symmetry is observed - also in the exactly counterrotating case -  with similar features to what we find: Reversals of the azimuthal velocity of the instantaneous flow are observed on the equatorial plane.  This fact can be interpreted as a phenomenon with characteristics much like the reversals of the vortex-like component represented by the second SVD mode. The laboratory flow can be divided into two toroidal cells around the two impellers, each one following the corresponding impeller, exactly as the mean flow of our model does. At the same time the fluid is driven towards the impellers and ejected to the cylindrical boundaries and the loop is closed with the return of the fluid along the walls and the equatorial plane. It turns out, however, that the flow presents two main configurations, each one breaking the symmetry around the equator (i.e., a rotation of $\pi$ around every axis which lies on the equatorial plane). Moreover, the flow configuration alternates between these two states, with spontaneous jumps from one state to the other. Each of these configurations presents a ``dominant cell'' (alternatively, the north or the south cell) characterized by a higher velocity and a larger spatial extent. In other words, the locus of points where the azimuthal velocity is zero does not lie on the equatorial plan, but encroaches on the northern (or in the mirror state, the southern) half-space. 

There are, however, distinct differences between these experimental results and our numerical results: In the experiment, the two configurations are relatively stable, i.e., the system can remain for relatively long periods in one of the two states before it undergoes a reversal. According to our simulations, the transition between the two states takes place not suddenly as in the experiment, but is better described as an oscillation (compare, e.g., Figure \ref{POD_time0002} with Figure 3a in \cite{navarra}). This difference probably emerges because of the smaller values of $Re$ in the simulations and because of the presence of an equatorial baffle in the implementation of the DYNAMO code (see \ref{role_baffle}). In fact, other experiments with the same cylindrical geometry, but with the presence of an inner equatorial ring with a strong impact on the turbulence (e.g., as described in \cite{Ravelet2005}), do \emph{not} observe a broken symmetry of this type \cite{navarra}.


\subsection{Role of time-stationary $u_2(\mathbf{r})$ in the dynamo process} \label{u2_role}

\begin{SCfigure} 
\begin{minipage}{.7\textwidth} 
\includegraphics[width=10.5cm, angle=0]{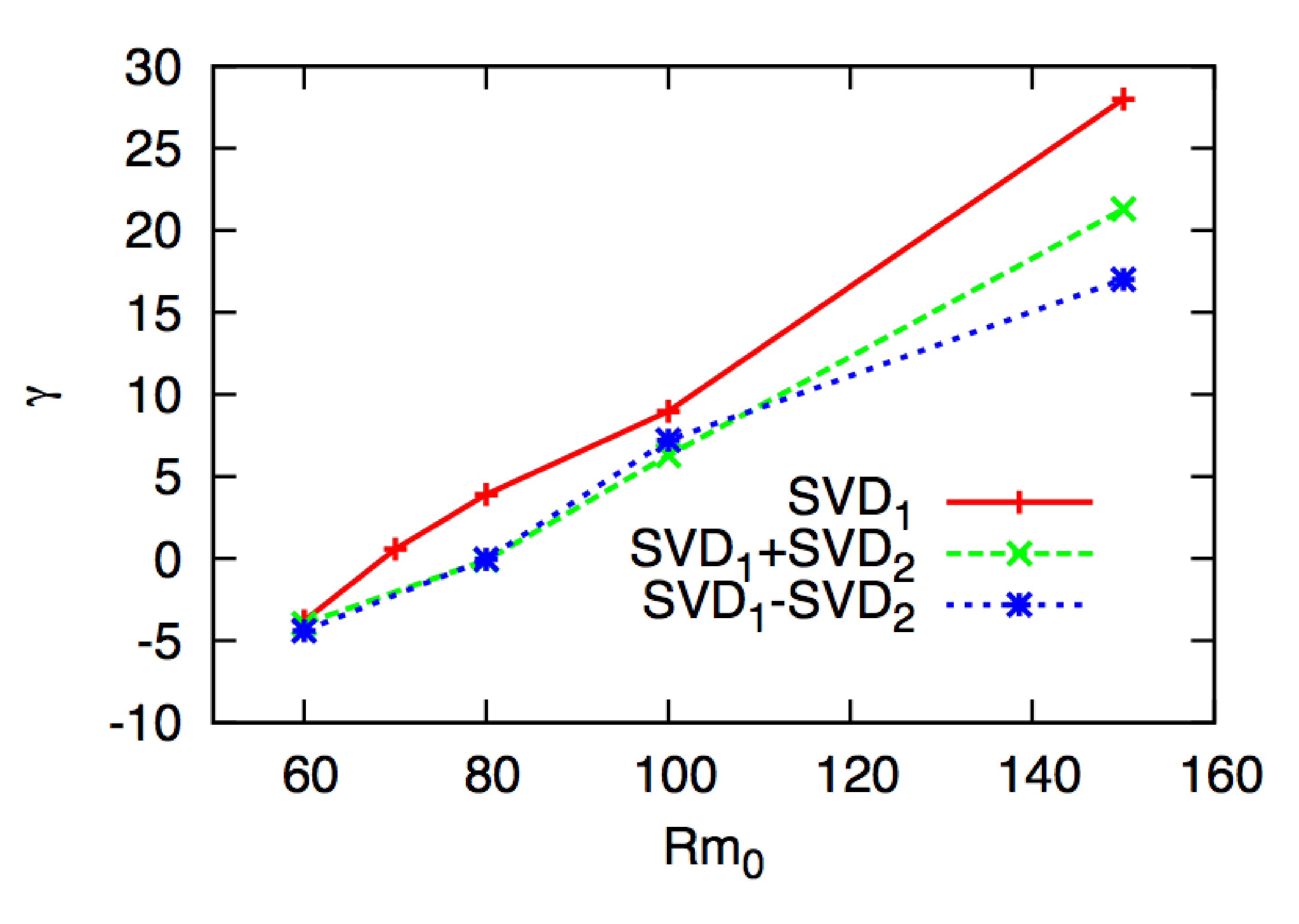} \centering \vspace{-1cm} \caption{\emph{Left}:~Growth rates $\gamma$ of the magnetic energy for the runs ``1'', ``1+2'', and ``1-2'', $Re_0~=~600$. A time-stationary $u_2(\mathbf{r})$ \emph{has a detrimental effect} on the kinematic dynamo threshold. $\gamma$~is scaled to the magnetic diffusion time, $\tau_\sigma=~\mu_0 \sigma L^2$ \cite{MePre}.} \label{impatto}
\end{minipage}
\end{SCfigure}

In order to identify the impact of $u_2(\mathbf{r})$ on the dynamo mechanism, we performed three kinds of time-stationary kinematic (i.e., the flow does not evolve and is decoupled from the magnetic field) simulations at $Re_0=600$ and different $Rm$. The first one uses as fluid flow only $u_1(\mathbf{r})$ (the dominant mode), weighted with the mean value $\langle\sigma_1 v_1(t)\rangle$ (see Fig. \ref{POD_time0001}). The second (third) run uses the superposition of $u_1(\mathbf{r})$ with $u_2(\mathbf{r})$, having weighted $u_2(\mathbf{r})$ with the maximum positive (minimum negative) value of the oscillating time trace $\sigma_2 v_2(t)$. The impact on the dynamo process can be summarized by comparing the growth rates of the magnetic energy and the critical magnetic Reynolds numbers (see Table \ref{impatto2} and Fig. \ref{impatto}, where the are runs indicated symbolically as ``1'', ``1+2'', ``1-2'', respectively).

As Fig. \ref{impatto} shows, part of the detrimental effect on the dynamo threshold at $Re_0=600$ can be ascribed to the presence of a time-stationary $u_2(\mathbf{r})$. At $Re_0=1100$ (see Table \ref{impatto2}), the difference in the growth rates for the ``1+2'' and ``1-2'' configurations can be understood in light of the fact that poorer statistics are accessible for this run.  The first SVD mode, $u_1(\mathbf{r})$, is noisier and less axisymmetric since the temporal sampling is shorter (the resolution is higher and the dataset was reduced in order to apply the SVD with the same CPU resources).  Adding and subtracting $u_1(\mathbf{r})$ and $u_2(\mathbf{r})$ produces different values of $\gamma$ because the noisy toroidal component of $u_1(\mathbf{r})$ tends to cancel out $u_2(\mathbf{r})$ in the ``1-2'' configuration. On the other hand, in the ``1+2''  configuration, $u_2(\mathbf{r})$ turns out to be detrimental. This last analysis does not take into consideration the role of the fluctuations in $u_2(\mathbf{r})$, which can also play a detrimental role, as suggested by the three-wave turbulent interaction picture in \cite{Kaplan2011, Kaplan2012}. The error bars of the exponents in Table \ref{impatto2} have been calculated via a standard linear regression of the data in a log-log plane using a least-squares method provided by GNUPLOT (see \cite{gnuplot}), which reports the standard deviation of the fitted curve $\tilde{\sigma}$ which is defined as the rms of the residuals. By using geometrical considerations on the growth rates of the magnetic energy (see, e.g., \cite{ReuterPhD}), it can be shown that the error on the slope of the curve (slope which corresponds to the exponents we are calculating) is $2\tilde{\sigma}/\Delta x$, where $\Delta x$ is the length of the range of the independent variable on which the fit is performed (in this case, the time interval over which the exponents are calculated).

\begin{table}[ht] 
\caption{Growth rates of the magnetic energy for the runs ``1'' and ``1+2'', $Re_0=1100$.} 
\centering \begin{tabular}{ c c c c} 
\hline\hline 
 & ``1'' & ``1+2''\\ \hline

$Rm_0$=100 & $5.64\pm0.013$ & $-3.8\pm0.4$ \\
$Rm_0$=150 & $27.4\pm0.4$ &$5.1\pm0.5$ \\ 


\hline

\hline
\end{tabular}

\label{impatto2}
\end{table}

\section{An interpretation of the $Rm_c(Re)$ curve} \label{curva}

One of the main results of \cite{Reuter2008} is the study of the stability curve $Rm_c(Re)$ (Fig. 6 in the paper). The curve represents the dependency of the critical magnetic Reynolds number of the dynamo instability on the fluid Reynolds number, i.e., the impact of turbulence on the onset of the dynamo. According to \cite{Reuter2008}, after a plateau in the laminar regime (at very small $Re$), the curve exhibits a quasi-linear behavior as $Re$ increases, demonstrating that the dynamo threshold becomes more difficult to reach when the system approaches the turbulent regime. Then, the stability curve shows a steep increase after this linear phase. Remarkably, $Rm_c$ reaches an absolute maximum, i.e., there is a value of $Re$ that makes the dynamo particularly difficult to obtain. Starting from $Re\simeq 1800$, the critical magnetic Reynolds number decreases and a saturation of the curve occurs at higher $Re$. In other words, regardless of the increase of turbulence in the fluid in this regime, the dynamo threshold remains constant and is not affected by a further increase of $Re$. These results can play an important role during the design phase of an experiment, especially if the curve stays constant at values of $Re$ that are higher than the ones explored by the simulations described in \cite{Reuter2008}. An interpretation of this flattening is given in \cite{Reuter2008} using scaling arguments within the framework of the Kolmogorov-Richardson phenomenology. 

Let us formulate another possible interpretation that can be given in the light of the results of the SVD analysis. As summarized in Table \ref{tabellare}, when $Re$ is increased, the relative energy content of the vortexes with a negative effect is enhanced by turbulence, whereas the energy in the mean flow would still depend only on the large scale forcing mechanism (and not on $Re$). In fact, the presence of $u_2(\mathbf{r})$ at $\mathrm{Re}_0=600$ enhances $\mathrm{Rm}_c$ by 20\%, whereas at $\mathrm{Re}_0=1100$ by 37\%. The picture suggests that turbulence, increasing the energy of $u_2$ and not of $u_1(\mathbf{r})$ - which is prescribed by the force - makes the dynamo instability more difficult to reach, because the energy is redirected into velocity modes with a negative impact on the magnetic field growth rate. This effect can hence explain the quasi-linear behavior that the $Rm_c(Re)$ curve shows at low $Re$, where an increase of $Re$ is converted into an increase of the critical magnetic Reynolds number. This same reasoning naturally leads to an explanation of the saturation of the $Rm_c(Re)$ curve at high $Re$ as well: As $Re_0$ increases, the detrimental modes that the SVD finds are not allowed to grow indefinitely without a specific force mechanism that drives them (i.e., a mechanism that is not present in the system under study, because the impellers drive an $s2t2$ flow), hence their negative impact on the dynamo is limited although the system becomes more turbulent.

\section{Strategies for turbulence control for dynamo purposes} \label{strategies}
In this section, we describe strategies designed in order to control large scale turbulence. The aim is to suggest practical actions to take in the experiments that can have a positive effect on the growth rates of the magnetic energy during dynamo action. We focus our attention on two different strategies. The first one (Section \ref{role_baffle}) is the control of shear layer turbulence, i.e., large scale eddies close to the boundaries with a detrimental effect on the dynamo. For this purpose, we model what was already installed in some experiments: a ring-shaped equatorial baffle, attached to the boundaries on the equatorial plane. As a second method (Section \ref{role_disc}), we implement the application of another baffle on the equatorial plane, disc-shaped, which was not used in the experiments and whose positive effects can be interpreted according to the results obtained via SVD.

\subsection{Role of an equatorial baffle}\label{role_baffle}
Hydrodynamic experiments \cite{Cortet2009,Ravelet2004} have shown that attaching an equatorial axisymmetric baffle to the inner surface of a cylindrical vessel (in the midplane of the cylinder) represents a suitable technique to control the fluctuation amplitudes of large scale flows. This strategy was used also for a dynamo experiment in cylindrical geometry \cite{Monchaux2007}. According to this study, such a baffle also has an effect on the mean flow, making the shear layer sharper around the midplane and, at the same time, reducing the turbulence intensity especially at low frequencies, although the flow remains strongly turbulent. The addition of such a baffle to a dynamo experiment was motivated by the results of previous numerical and experimental investigations that explored the influence of turbulence (especially at large scales) on the dynamo threshold, and of the transverse motion of the shear layer across the midplane.  For example, the addition of large scale noise to the Taylor-Green mean flow increases the dynamo threshold, as shown numerically in \cite{Laval2006}.  In addition, fluctuating motion of eddies of the Roberts flow increase the threshold \cite{Petrelis2006}. In \cite{Volk2006}, it is reported that the magnetic induction (due to an externally applied field) on a gallium flow depends significantly on the large scale flow fluctuations. 
All the results described so far suggest that the large scale motion and its fluctuations are of crucial importance for the creation of a dynamo in the MDE, although there is no concrete theoretical prediction about this issue. Although the experiments mentioned above have a slightly different geometry from the MDE, fluctuating large scale eddies in the MDE can create a great difference between the instantaneous flow and the mean flow \cite{Nornberg2006a}, creating a detrimental effect for the magnetic field growth, although the mean flow - if stationary - is optimal (according to \cite{Dudley1989}). In the MDE, with the application of an external axial magnetic field, it is possible to observe an intermittent behavior of the magnetic field (whose fastest-growing eigenmode can alternate growing and decaying phases, see \cite{Nornberg2006a}). This result is seen as an effect of the departure of the instantaneous field from the mean field. In \cite{Kaplan2011}, it has been shown that such a baffle can successfully reduce global turbulent resistivity in the MDE.

\subsubsection*{Implementation of the equatorial baffle.}
 The baffle effect consists of a modification of the computed velocity field by suppressing the vertical component of the flow. This suppression is applied at every time step in the spatial domain and then transferred to the spectral domain by a spatial-to-spectral transform. The flow modification takes place close to the boundaries, acting as a filter on the predicted velocity field. More precisely, the $z$ component of the field was damped according to the following operation
 
 \beq\mathbf{v}(r,\theta,\phi)\rightarrow\mathbf{v}(r,\theta,\phi)-v_z f(r,\theta,\phi)\hat{\mathbf{z}},\eeq where the function $f$ carries the spatial information of the damping mechanism. In order to avoid abrupt damping effects and steep gradients in the simulations (that can be a source of numerical instabilities and Gibbs phenomena), the function $f$ employs Gaussian functions to describe the transition between the region where the damping effect is present and the region where it is not. The functional form of $f$ was chosen as
 
 \beq f(r,\theta,\phi)=\exp\{{-\frac{r^2\cos^2{\theta}}{2\sigma^2_z}}\} f_{ring}(r,\theta,\phi),\label{impl_filter}\eeq
 
 where, in turn,
 \beq
 f_{ring} =
\left\{
	\begin{array}{ll}
		\exp\{-[(r\sin\theta-\rho_{ring})^2]/(2\sigma^2_r)\}   &\quad r\sin{\theta} < \rho_{ring} \\
		1 &\quad r\sin{\theta} > \rho_{ring} 
	\end{array}
\right.
 \eeq
 The interpretation of the adjustable parameters $\sigma_z$, $\sigma_r$, and $\rho_{ring}$ is as follows. The first parameter, $\sigma_z$, is a measure of the thickness of the baffle; $\sigma_r$ represents a damping parameter along the radial direction; $\rho_{ring}$ is a measure of the difference of the radius of the sphere and of the radius of the baffle (in the experiment, it is 8 cm).
  
\subsection{Role of a disc-shaped equatorial baffle between the impellers}\label{role_disc}

Another strategy for suppressing detrimental dynamics follows naturally from the results of the SVD analysis described in this paper.  We implemented in the code the effect of a discoidal baffle centered on the equatorial plane (i.e., lying between the impellers), in order to separate the dynamics in the two hemispheres, avoiding a strong poloidal circulation near the axis, and facilitating the separation of the dynamics of the toroidal circulation in the two hemispheres. In the light of the SVD results, the underlying idea is that the disc can influence $u_2(\mathbf{r})$ and hence facilitate the dynamo.

\subsubsection*{Implementation of the disc-shaped baffle.}
 The action of this disc consists in a modification of the computed velocity similar to the one used for the ring-shaped baffle. The field is modified in a flat circular region lying on the equatorial plane where a spatially smooth suppression of the vertical component of the flow is performed at every time step. This suppression acts as a filter on the predicted velocity field. The $z$ component of the field is damped according to Equation \ref{impl_filter}, using another $f$ function in order to apply the damping mechanism in the center of the equator. Again, the function $f$ employs Gaussian functions to describe the spatial behavior of the disc in order to avoid numerical instabilities (for this reason, the disc is approximated by a flat ellipsoid). Thus, the functional form of $f$ for the disc is
 
 \begin{equation} f(r,\theta)=\exp\Big[{-\frac{r^2\cos^2{\theta}}{2\sigma^2_z}}-\frac{r^2 \sin^2{\theta}}{2 \rho^2_d}\Big] .\end{equation} The interpretation of the adjustable parameters $\sigma_z$, and $\rho_{d}$ is straightforward: $\sigma_z$ is a measure of the thickness of the disc; $\rho_d$ represents the disc radius, whose value was chosen in order to cover spatially at least the region of the equatorial plane through which $u_2(\mathbf{r})$ flows, as Figures \ref{2POD_pol_new} and \ref{2POD_tor_new} suggest.
 
 \subsection{Results}\label{impl_res}
Kinematic simulations (i.e., neglecting the Lorentz force term in the Navier-Stokes equations) show that the growth rate $\gamma$ of the magnetic energy is enhanced by both the baffle and the disc. Nevertheless, the baffle effect is not as strong as the disc effect. Table \ref{tabellare2} shows the effect of the disc on the default growth rate $\gamma$. We indicate with $\gamma_r$ the growth rate of the magnetic energy in presence of the baffle, $\gamma_d$ in presence of the disc. It should be noted that a larger $\rho_d$ is needed to have a non negligible effect at $Re_0=1000$, confirming the necessity of a stricter control of large scale turbulence at higher $Re_0$. The results are discussed in section \ref{results_inter}.

\begin{table}[ht] 
\caption{MHD simulations of dynamos with an implemented equatorial disc in the center of the equatorial plane. $\gamma_d$ is the growth rate (the time is scaled to the resistive diffusion time, $\tau_\sigma=\mu_0 \sigma L^2$) of the magnetic energy in the presence of the disk; $\gamma_r$ in the presence of the ring; $\gamma_0$ is the default growth rate, i.e., without any baffle. As stated in the text, the initial choice of $\rho_d$ has been motivated by considering the curves shown in Figs \ref{2POD_pol_new}-\ref{2POD_tor_new}.} 
\centering \begin{tabular}{ c c c c c c c} 
\hline\hline 
 $Re_0$ & $Rm_0$ & $l_{max}$ & $\rho_d$ & $\gamma_r$ & $\gamma_d$ & $\gamma_0$   \\ \hline
300 & 80 & 30 & 0.28 &7.62 &18.8 & 4 \\
300 & 100 & 30 & 0.28 &15.5 &30.8 & 9.8\\
300 & 150 & 30 & 0.28 & 31.2 &58.6 & 24\\
300 & 250 & 30 & 0.28 &52 &109 & 39\\
600 & 250 & 52 & 0.28 &28.4 &45.4 & 22.6 \\
600 & 250 & 72 & 0.28 & 29.9 & 50.4 & 22.6 \\
1000 & 300 & 52 & 0.65 &10.4 & 18.4 & 10.4 \\

\hline
\label{tabellare2}
 \end{tabular}

\end{table}

\section{Interpretation of the results}\label{results_inter}

In \cite{Dudley1989}, Dudley and James studied the kinematic dynamo threshold of several flows. It turned out that simple flows (like the \emph{s2t2}, the \emph{s2t1} or \emph{s1t1}) can trigger a dynamo field under specific circumstances. In particular, the flow \emph{s1t1} investigated in \cite{Dudley1989} is \begin{equation}\mathbf{u}=\mathbf{t}^0_1+\epsilon \mathbf{s}^0_{1},\end{equation} with Bullard-Gellman stream functions \begin{equation}s_1^0(r)=t_1^0(r)=r\sin{(r\pi)},\end{equation} and $r\in [0,1]$. These modes are similar to the corresponding ones of the second SVD mode, although they are not peaked near the axis as in our case. If $\epsilon=0.17$ and $Rm>155$, this flow turns out to sustain a kinematic dynamo.

The fact that the growth rate of the magnetic energy is negative for a whole range of low $Rm$, where \emph{s2t2} is already a dynamo, can explain the reason why the combinations $\mathbf{u}_1\pm\mathbf{u}_2$ have lower growth rates than the single $\mathbf{u}_1$. Following the discussion in \cite{delatorre2007} (where the influence of time dependent flows on the kinematic dynamo threshold in v\'an K\'arman flows was numerically investigated), the magnetic field responds to the instantaneous velocity field, oscillating between instantaneous solutions of the induction equation, and resulting in average growth rates that are smaller than those produced by the optimal configuration (in our case the \emph{s2t2} flow). A possible three-wave interaction between the velocity modes and magnetic modes could play a crucial role determining into which magnetic modes the kinetic energy flows.  It is expected that one role of the $u_2(\mathbf{r})$ mode is to extract energy from the magnetic field dipolar dynamo mode and put it into other magnetic modes that do not grow, thereby reducing the magnetic energy growth rate of the system. This idea will be tested by future investigations.

The two strategies described in section \ref{strategies} turn out to have a positive effect on the dynamo growth rate. In particular, by comparing the growth rate $\gamma_r$ in the presence of the ring, the growth rate $\gamma_d$ in the presence of the disc, and the growth rate $\gamma_0$ without any baffle, it is observed that the disc has the strongest impact on $\gamma$. This result is consistent with the SVD analysis presented in this study: most of the turbulent detrimental action takes place along the axis, and not exclusively near the boundaries. This suggests a strategy for facilitating the dynamo instability: it is expected that more robust dynamo activity could be achieved in experimental configurations by controlling the symmetry of the counter-rotating flow in the equatorial region, avoiding a strong poloidal circulation near the axis, and ensuring the separation of the dynamics of the toroidal circulation in the two hemispheres.          

\section{Summary}

In this work, we presented computational results with direct application to the generation of magnetic fields in liquid metal experiments.  Dynamos are often not observed in laboratory experiments at Reynolds numbers where they are theoretically expected, and understanding the reasons for this discrepancy is fundamental in order to implement new experimental measures designed to overcome this problem.  The key result of this work was the identification, by the use of SVD, of turbulent dynamics that constitute an obstacle for dynamo activity.
These results have produced new insights into the hydrodynamics of the flows that are important for producing the dynamo mechanism in experimental configurations.  

A detailed analysis of the fluid dynamics is the key issue, and the Singular Value Decomposition constitutes a tool for determining the eigenmodes that best suit the problem. Furthermore, SVD identifies the temporal behavior of each of these modes, and quantitatively defines their relative importance. With these techniques, we identified specific features of the turbulent flow that - in combination with the background flow - suppress magnetic field generation. In other words, we proposed an easily classifiable flow structure that is detrimental to the dynamo mechanism, and provided an additional support for large scale turbulence suppressing the dynamos rather than small scale.

We also studied the effect of two strategies designed to control large scale fluctuations in our numerical model: The implementation of ring-shaped and disc-shaped baffles. The ring has already been used in experiments, and was shown to reduce turbulent resistivity in experiments, and to slightly increase the growth rates in the simulations. This study indicates that the disc is more effective at facilitating dynamo activity, as it more strongly increases the growth rates of the magnetic energy.  This result can also be useful for experiments with similar topologies, since it constitutes a concrete step to take in order to overcome possible suppression mechanisms generated by fluctuations that break the so called $\pi$-symmetry of the system (i.e., a rotation of $\pi$ around every axis that lies on the equatorial plane). 

Future work could be focused on the details of the effects on the magnetic field of the secondary dynamics identified in this study. A three-wave interaction (between velocity modes and growing/damped magnetic modes) is playing an important role, and the SVD could be useful for elucidating these dynamics, e.g., by applying the SVD to the magnetic field and investigating correlations with the most important $u_i(\mathbf{r})$'s. In fact, we expect that one role of $u_2(\mathbf{r})$ is to extract energy from the magnetic field dipolar dynamo mode and put it into other damped magnetic modes.

\ack
The computations were performed on the BOB and TOK clusters hosted at the Garching Computing Center (RZG) of the Max Planck Society. A.L. and F.J. gratefully acknowledge financial support of DFG (project on ``Simulations and theory of mechanically driven, turbulent dynamos in spherical geometry''). The research leading to these results has received funding from the European Research Council under the European Union's Seventh Framework Programme (FP7/2007-2013) / ERC grant agreement n$^{\circ}$ 277870.

\section*{References}
\bibliographystyle{unsrt}
 \bibliography{POD}

\end{document}